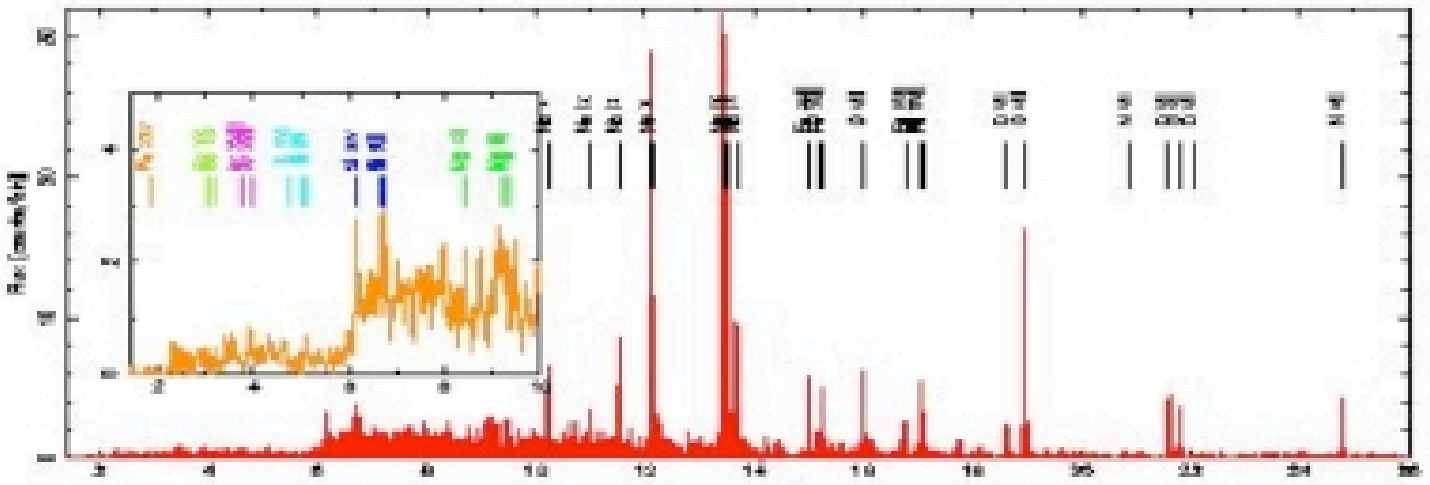

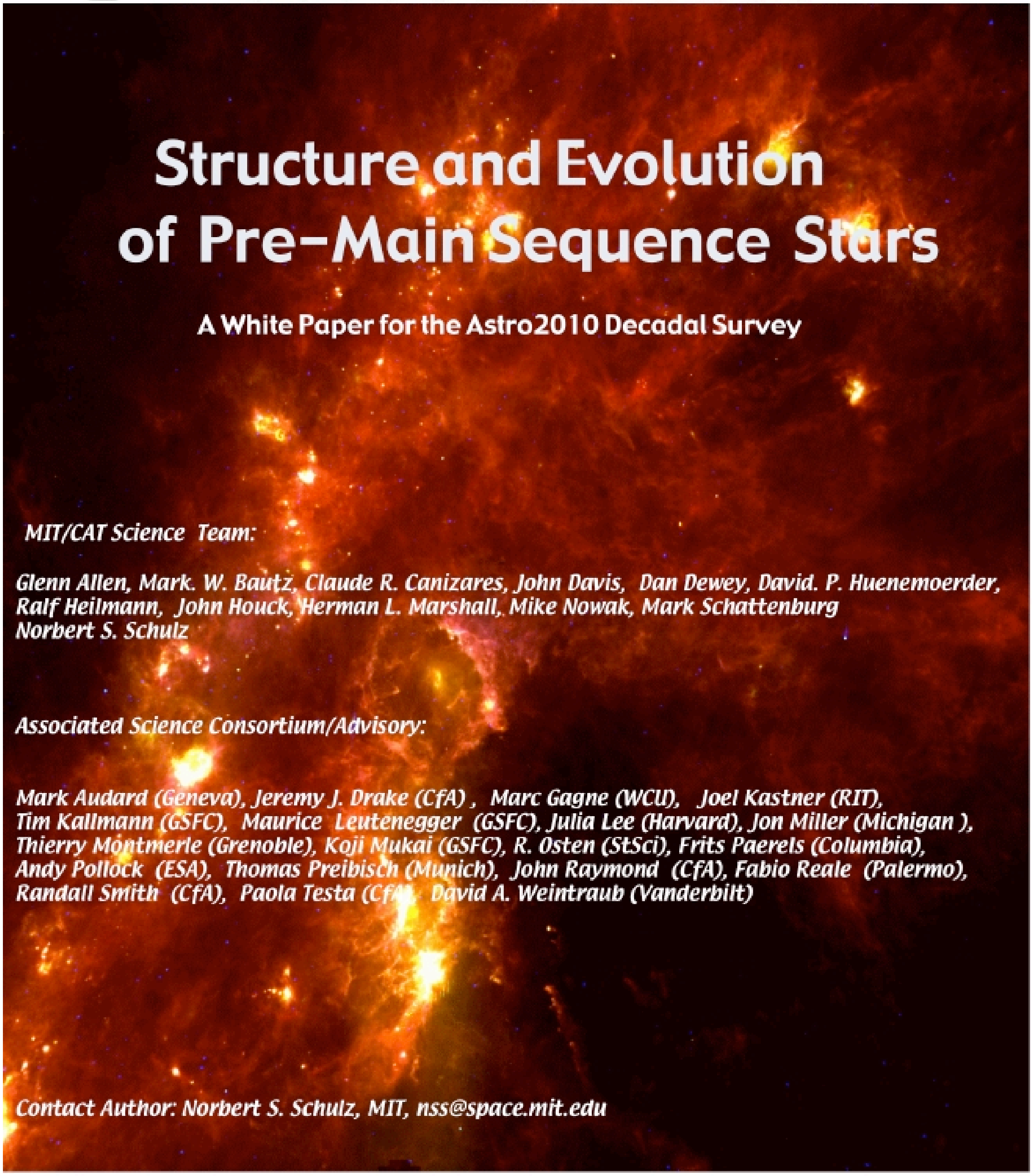

# Structure and Evolution of Pre-Main Sequence Stars

A White Paper for the Astro2010 Decadal Survey

*MIT/CAT Science Team:*

*Glenn Allen, Mark. W. Bautz, Claude R. Canizares, John Davis, Dan Dewey, David. P. Huenemoerder, Ralf Heilmann, John Houck, Herman L. Marshall, Mike Nowak, Mark Schattenburg Norbert S. Schulz*

*Associated Science Consortium/Advisory:*

*Mark Audard (Geneva), Jeremy J. Drake (CfA), Marc Gagne (WCU), Joel Kastner (RIT), Tim Kallmann (GSFC), Maurice Leutenegger (GSFC), Julia Lee (Harvard), Jon Miller (Michigan), Thierry Montmerle (Grenoble), Koji Mukai (GSFC), R. Osten (StSci), Frits Paerels (Columbia), Andy Pollock (ESA), Thomas Preibisch (Munich), John Raymond (CfA), Fabio Reale (Palermo), Randall Smith (CfA), Paola Testa (CfA), David A. Weintraub (Vanderbilt)*

*Contact Author: Norbert S. Schulz, MIT, nss@space.mit.edu*

***Abstract***

*Low-mass pre-main sequence (PMS) stars are strong and variable X-ray emitters, as has been well established by EINSTEIN and ROSAT observatories. It was originally believed that this emission was of thermal nature and primarily originated from coronal activity (magnetically confined loops, in analogy with Solar activity) on contracting young stars. Broadband spectral analysis showed that the emission was not isothermal and that elemental abundances were non-Solar. The resolving power of the Chandra and XMM X-ray gratings spectrometers have provided the first, tantalizing details concerning the physical conditions such as temperatures, densities, and abundances that characterize the X-ray emitting regions of young stars. As a result, we now know that phenomena leading to X-ray emission are much more diverse; we have strong evidence that significant X-ray emission can originate from accretion processes in addition to more complex coronal structures. Key diagnostics for probing the physical structure are the emissen line fluxes of elements from carbon through nickel, particularly the helium- and hydrogen-like series and the iron L series, and the dynamical effects as determined from line positions, shapes, and variability.*

*Over the last 20 years X-ray imaging spectroscopy with medium spectral resolution has observed all of the important nearby regions and high resolution spectroscopy will bring further progress in this area. The high resolution spectrometers on Chandra and XMM, however, simply do not have the effective area to measure diagnostic lines for a large number of PMS stars over required to answer global questions such as: how does magnetic activity in PMS stars differ from that of main sequence stars, how do they evolve, what determines the population structure and activity in stellar clusters, and how does the activity influence the evolution of protostellar disks. Highly resolved (R>3000) X-ray spectroscopy at orders of magnitude greater efficiency than currently available will provide major advances in answering these questions. This requires the ability to resolve the key diagnostic emission lines with a precision of better than 100 km/s.*

*Only by obtaining X-ray spectra of young stars in nearby star forming regions, young embedded clusters in the Milky Way, and in star forming regions in nearby galaxies can we conclude studies begun with imaging-spectroscopic surveys. The detailed spectra that do exist are limited to a few of the brightest (and possibly anomalous) sources. High resolution X-ray spectra are the only means. With such capability, we can obtain X-ray spectra of complete samples of protostars (age << 1 Myr) in embedded clusters, of PMS stars in star-forming clouds (age 1-10 Myr), and of more "evolved" (age 10-100 Myr), widely distributed young stars, and we can do so throughout the Milky Way. Such a high-resolution spectroscopic database is absolutely necessary if we are to properly interpret the large and growing volume of imaging spectroscopic data collected for young star clusters by Chandra and XMM. The resulting high-resolution spectra are the only means to separate contributions of accretion from that of coronal magnetic activity. They are essential for understanding both the operation of magnetic dynamos in rapidly rotating stars and accretion through a magnetosphere.*

## *1. Introduction – X-Rays from Young Stars*

High resolution X-ray spectra of young stars are important to understand processes that shaped the appearance of our Sun today as well as the impact of these processes on the evolution of the protosolar disk and planets. Until the launch of ROSAT and the Chandra the study of young stars of all masses and sizes was entirely restricted to long wavelength observations, which used optical and IR observations mostly to characterize the properties of circumstellar gas and dust. From these studies Lada et al. 1987 indentified three classes of young stellar systems based on their infrared spectral energy distributions (SEDs), where class I sources are evolved protostars with large IR excesses from circumstellar envelopes, class II sources are accreting pre-main sequence stars with moderate IR excesses and strong $H_\alpha$ line emissions, Class III sources are mainly contracting PMS stars with some IR excess from a residual protostellar disk without an inner accretion disk and thus weak $H_\alpha$ line emissions. The former are also known as Classical T Tauri Stars (CTTS), the latter as Weak Lined T Tauri Stars (WTTS). Later a Class 0 was added to depict highly embedded unevolved protostellar systems.

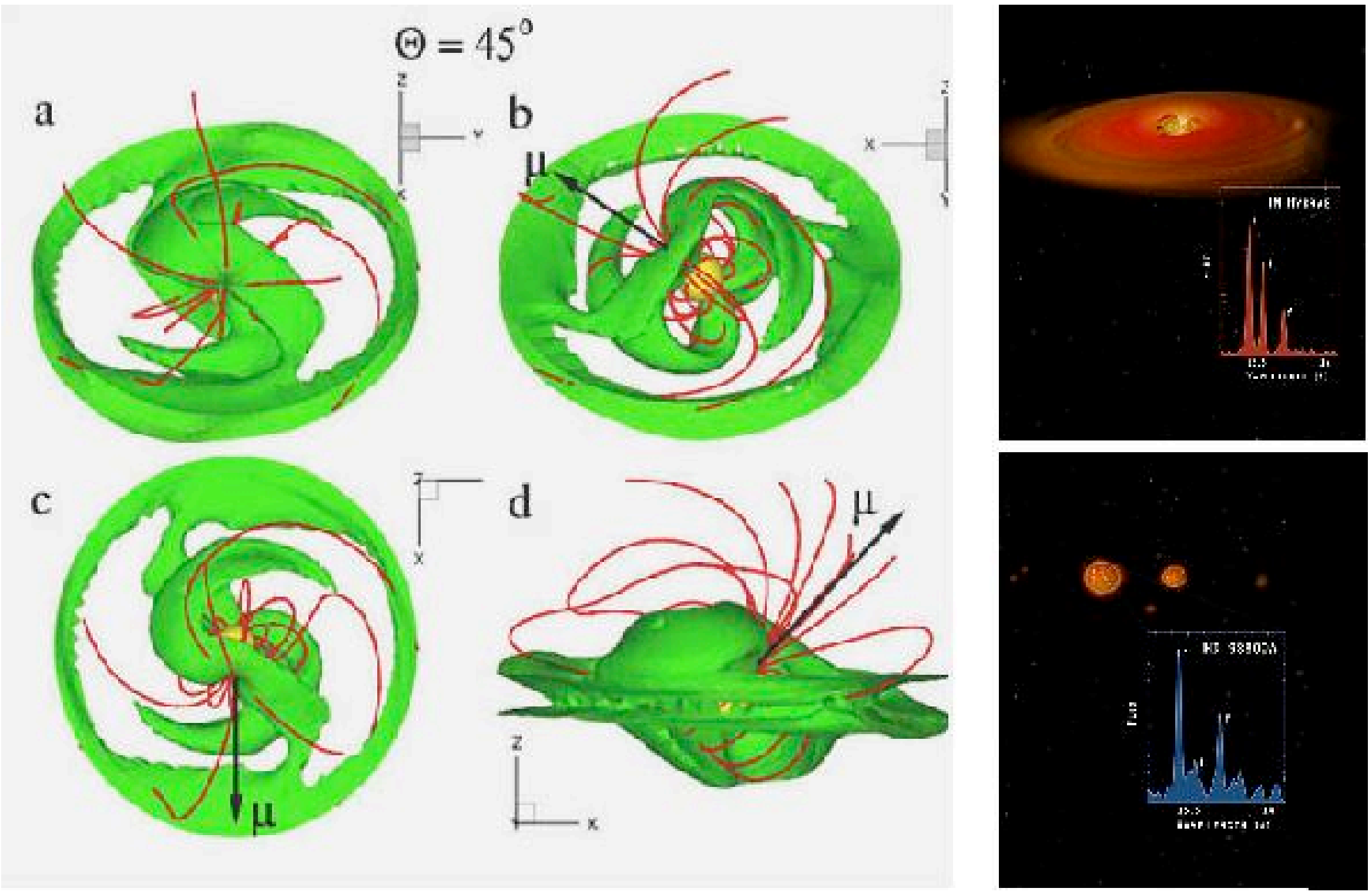

***Figure 1:*** *(Left) Various views of recent simulations of magnetospheric accretion by Romanova et al. (2007) in a young CTTS. (Right) Highly resolved Ne IX triplets indicate X-rays dominated by accretion in the CTTS TW Hya and by coronal emission in the WTTS TV Crt.*

The classification scheme proved successful in optical and IR studies which lead to a standard picture of evolutionary phases with many global details. X-ray studies of nearby star forming regions predominantly with EINSTEIN, ROSAT and Chandra produced a good congruency between IR and X-Rays in terms of detections but not in terms of SEDs - WTTS seemed to emit stronger X-rays than CTTS. This suggests that X-rays trace magnetic properties of the evolving star, while the IR traces reprocessing within circumstellar material.

High resolution spectra of members of the nearby TW Hya association provided by Chandra showed that X-ray emission in the early phases of T Tauri stars is still significantly affected by accretion shocks rather than purely coronal emissions (Kastner et al. 2002, 2004; Fig. 1, right). Simulations show that star and protoplanetary disk are intertwined and accretion proceeds via a complex structure of magnetic field lines (Figure 1, left). The following questions have been raised during the Chandra Workshop "Star Formation in the Era of Three Great Observatories" involving the Spitzer. Hubble, and Chandra Observatories (Wolk et al. 2006):

- *What are the main drivers in the evolution of protostellar disks?*
- *What is the population structure of a star forming association?*
- *How do the dynamo processes differ in PMS stars as compared to MS stars?*

These are fundament questions where studies of high resolution X-ray spectra can provide answers to. They were addressed to the community for the remainder of these missions but also for future observatories. Below we outline three projects which specifically require high spectral resolution in the X-band from 200 eV to 1.5 keV (8 – 50 Angstrom) and 6.2 – 7.0 keV (1.7 – 2.0 A) to provide meaningful solutions to some of these questions. More IXO contributions to star formation appear in the white papers by Feigelson et al. and Osten et al. for this decadal survey.

### 2. *Fundamental Processes: X-ray Accretion Shocks and Coronal Evolution*

Emission and absorption lines are the most powerful tools to extract physical parameters of astrophysical plasmas. In the case of stars with magnetic activity we are specifically interested in line emission from magnetically confined coronal plasmas. Since Chandra and XMM we now know that there are at least two major contributing mechanisms in young pre-main sequence stars (Figure 1).

One mechanism describes plasma flows from active disk accretion along magnetic field lines onto the young stellar surface (see Figure 1, left) which produces accretion shocks of moderate temperatures ( < 3-5 MK). Plasma is compressed to densities well beyond $10^{11}$ $cm^{-3}$, so that the triplet lines of the He-like ions from CV to Ng XI become sensitive density diagnostics (Ness et al. 2002, Testa et al. 2004). The Chandra and XMM spectra of the brightest PMS sources have permitted the first attempts to obtain densities (Kastner et al. 2002, 2004; Schmitt et al. 2005). The fact that accretion shocks appear in CTTS spectra came somewhat of a surprise as they had been predicted to radiate in the UV band at most, not in X-rays (Hartmann 2001). This fact presents scientists with an opportunity to study star-disk interactions and magnetic configurations in X-rays (Figure 2, left) and to obtain an independent measure of accretion rates (Guenther et al. 2007). X-rays drive chemical processes in circumstellar disks, they produce ionization that may enable the Magneto-Rotational instability to drive accretion in the upper layers of the disk (e.g. Glassgold et al. 1997). The X-rays are instrumental in driving significant photo-evaporative flows from the disk, either through direct X-ray heating (Ercolano et al 2008) or in combination with FUV-EUV irradiation (Gorti & Hollenbach 2009). Mass loss rates through photo-evaporation can be comparable to accretion rates and are important for disk and protoplanetary evolution. The efficiency of an X-ray driven photoevaporative wind mechanism is still a matter of dispute in literature. While Alexander, Clarke, & Pringle (2004) conclude that X-rays have no significant effect, Glassgold et al. (2004) argue for disk heating and changes in chemistry.

The second fundamental process is the magnetic dynamo responsible for coronal activity. Resolving the coronal emission from PMS stars of various ages will enable to study the evolution of the young stars coronal activity. We know now it is likely that in young CTTS X-rays are still produced by accretion. Eventually coronal emission overtakes accretion and coronal activity become dominant during the stars's contraction phase. The strong coronal activity should evolve into the moderate activity our Sun displays today scaled by rotation. Not much is known about what causes coronal activity in evolving PMS stars. Various forms of internal turbulent dynamos (Figure 2, right) have been proposed in contrast to the solar shear interface dynamo responsible for the X-ray activity shown in the SOHO inset image. IXO will be able to perform Doppler analysis of CTTS rotation rates for at least the rapid rotators exhibiting 50 – 100 km $s^{-1}$ splittings offering a spectroscopic view at the structure of the magnetosphere (Figure 3).

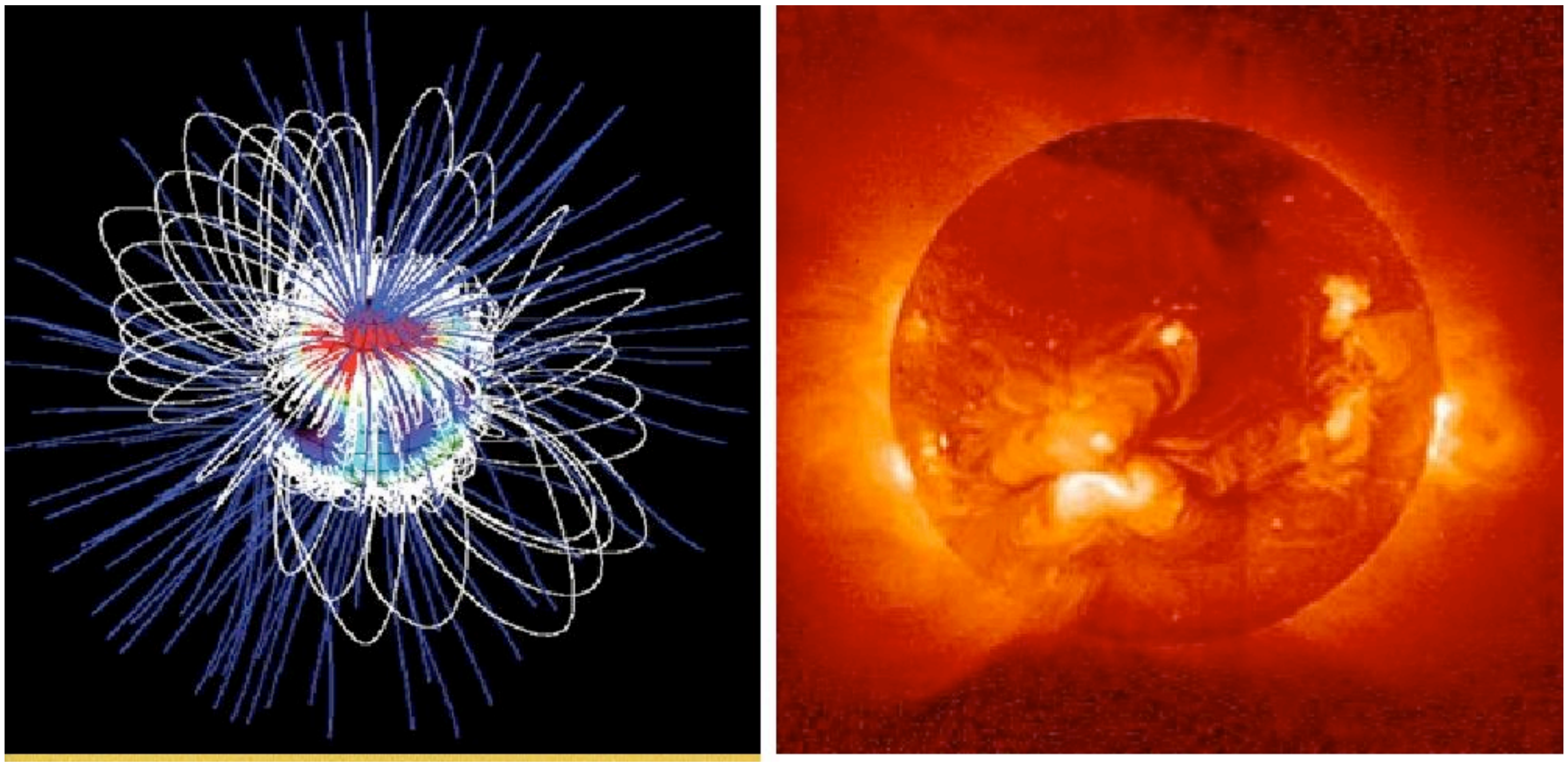

***Figure 2:*** *(Left) Magnetospheric topology of the CTTS V2129 Oph as derived from an extrapolations of a surface magnetic topology reconstruction (from Donati et al. 2007) which may be the result of internal turbulent dynamo activity. (Right) Soho X-ray image of the Sun.*

The nature of the underlying dynamo can also studied though stellar X-ray flaring activity. On the main-sequence it is known that in rapidly rotating stars an $\alpha^2$ dynamo takes effect. How this may relate to PMS evolution remains to be determined, maybe through measurements of turbulent velocities. Clearly, the connection between the stellar field and the disk produces magnetic stresses that would be relieved by small scale or large scale flaring. Stellar flares on cool stars are a ubiquitous phenomenon in the X-ray spectral region. Studies of the Sun reveal that this most dynamic aspect of magnetic activity is a primary source of plasma heating. Solar flares are known to be a manifestation of the reconnection of magnetic loops, accompanied by particle beams, chromospheric evaporation, rapid bulk flows or mass ejection, and heating of plasma confined in loops. For young stars this implies that loops and reconnection can affect the angular momentum evolution of proto-stellar disks. Significantly increased X-ray sensitivity and resolution would routinely provide powerful new diagnostics - Fe

K fluorescence, line profiles, and line shifts (see Figure 3) - probe non-equilibrium plasma states through flare evolution, and constrain the effects of X-ray flare irradiation on circumstellar disks.

In order to pursue these science goals we would like to employ several diagnostics based on accurate knowledge of X-ray line emissivities. Observations of X-ray lines provide reliable plasma parameters, in particular temperatures, densities, and elemental abundances. For density diagnostics in the X-ray band we require sensitivity for densities between $10^{11}$ $cm^{-3}$ and a few $10^{13}$ $cm^{-3}$ which are most effectively provided by the He-like triplets of O VII, Ne IX, and Mg XI. Specifically the latter two ions require very high spectral resolution in order to effectively separate contaminating line blends. This is illustrated in Figure 3 for the wavelength band between 13.5 and 13.9 Angstrom, which hosts the triplet components of Ne IX (Ness et al. 2003), for a resolution of R = 3000 (top) and R = 300 (bottom). Amidst a forest of transitions from Fe ions, the Ne IX lines can only be fully separated at high resolution. The turbulence diagnostics are similarly stringent. It has been shown in active coronal sources such as II Peg (Huenemoerder et al. 2001) and AR Lac (Huenemoerder et al. 2003) that X-ray emission lines observed with Chandra are usually unresolved. This is expected , as models predict line widths from coronal loop emission to be only a few tens of km $s^{-1}$ , which in the X-ray band below 12 Angstrom requires resolving powers of 3000 – 5000.

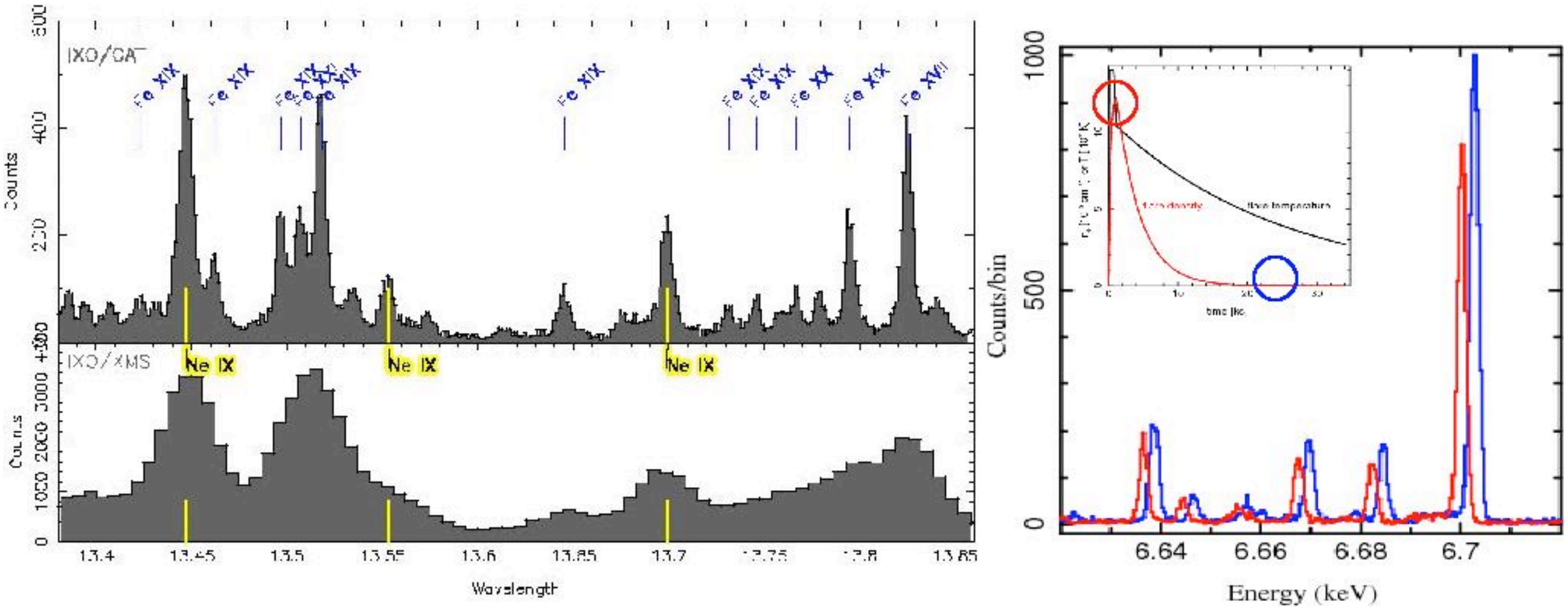

***Figure 3:*** *(left) The X-ray band pass of the He-like triplet of Ne IX. The simulation shows a comparison of the low resolution (R = 300, bottom) and the high resolution (R = 3000, top) case. (right) An adopted temperature and density evolution of a flare event. During the initial heat pulse (red circle and lines) we assumed a bulk velocity of -100 km/s and an emission measure reconstruction as described by Reale et al. (2004). The flow in the short thermal pulse is e.g. seen in the Fe XXV lines (blue is for t=0-2ks, and red is 2-4 ks).*

Increased X-ray sensitivity and resolution (> 1500) will routinely provide powerful new diagnostics through extensive stellar surveys, probe non-equilibrium plasma states through flare evolution, and constrain the effects of X-ray flare irradiation on proto- and circumstellar disks. A theoretical foundation to diagnose the rise as well as decay of a single-loop flare is provided by Reale 2007.

### *3. Young Stars in Nearby Star Forming Regions and Beyond*

Chandra and XMM together observed about two dozen of the brightest PMS stars with adequate X-ray line statistics to allow meaningful plasma diagnostics. Examples include the brightest stars of the TW Hya association (Kastner et al. 2002, 2004, Huenemoerder et al. 2007), BP Tau (Schmitt et al. 2005), V4046 Sagittari (Guenther et al. 2006), and some ongoing activities related to the Chandra HETG Orion Legacy Project (Schulz et al. 2008). In order to pursue the science above we need survey capability to obtain to spectra of hundreds of resolvable nearby star forming regions within 20 to 100 pc including the Taurus-Auriga complex, the Chameleontis, Ophiuchus, and Lupus regions, to name a few examples (Zuckerman and Song 2004, Torres et al. 2009). In these regions, future X-ray observatories such as the planned International X-Ray Observatory (IXO) will have enough spatial resolution to resolve and efficiently observe hundreds of PMS stars. It works to our advantage that most of these stars are well studied in longer wavelength bands and are also focus of more future studies with upcoming observatories. In other words, important basic parameters such as age, effective temperature, and bolometric magnitudes will be available to interprete the X-ray results.

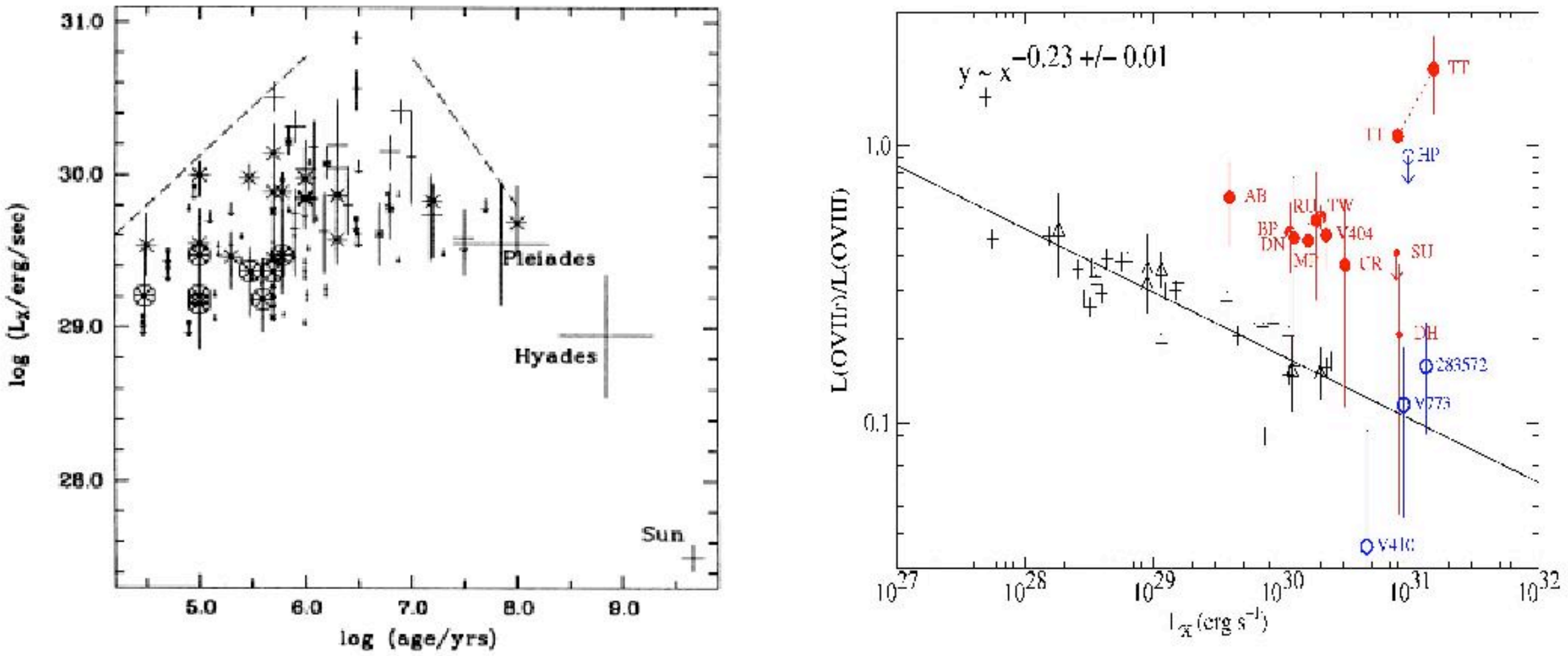

***Figure 4:*** *(Left) X-Ray luminosities vs age for T Tauri stars in the Taurus region in relation with cluster ages (Hyades and Plejades) near the main sequence as well as the Sun (from Neuhaeuser et al. 1995). (Right) Ratio between the OVII resonance line and OVIII Lya line luminosities vs. total X-ray luminosity $L_x$. The solid line is a power-law fit to the main sequence stars with $L_x > 10^{27}$ erg/s (from Guedel and Telleschi 2007).*

The ROSAT All Sky Survey (RASS) detected several thousand PMS stars. And although Chandra not only followed up on many of them but also resolved many more in embedded clusters with imaging spectroscopy, all but the brightest few tens are beyond the reach for the Chandra and XMM gratings. The Upper Scorpius region is one of the many which includes a variety of T Tauri and post-T Tauri stars where the clustered stars also coincide with brighter X-ray sources, while the more dispersed populations are X-ray quiet (Preibisch et al. 1998). The apparent age dependence of X-ray luminosity is now well established (see Figure 4, left) and provides evidence for the complexity of coronal evolution in PMS stars. A recent survey involving higher X-ray resolution has been the XEST survey studying soft excesses in CTTS with the XMM-RGS at resolving powers of around 500 (Guedel et al. 2007). At these resolving powers O VII and O VIII line fluxes can be extracted as long as the O VII widths remain below well below 1000 km $s^{-1}$.

Figure 4 (right) hightlights the detection of excess cool (1-3 MK) material in the O VII line in CTTS. Fully resolved X-ray line emission at spectral resolving powers of at least 1500 to 3000 will be the next logical step to continue what ROSAT, Chandra, and XMM have started . This involves the determination of complete abundance and temperature distributions, detailed accretion geometries, and coronal properties with the 1efficiency standard astronomical surveys require. The only instruments that allow such studies in the future are currently envisioned for IXO in the next decade and for Gen-X in 20 – 30 years. For example, IXO alone can efficiently observe high resolution spectra of several 1000 nearby PMS stars with X-ray fluxes down to between $10^{-12}$ to $10^{-13}$ erg $cm^{-2}$ $s^{-1}$ (~ $10^{-14}$ erg $cm^{-2}$ $s^{-1}$ in the 0.2 – 10 keV bandpass) and produce age-luminosity tracks for accretion and magnetic processes separately.

There is a possibility that these surveys can also be extended beyond nearby regions and launched throughout not only our own Galaxy, but also into our neighbors such as the LMC, SMC, and M 31. Lada and Lada (2003) listed about 150 embedded stellar clusters within a radius of 3 kpc of the Sun. Though these cannot be spatially resolved, line ratios from line fluxes of cummulative spectra alone can lead to statistical surveys among many clusters, specifically in terms of abun-dance ratios such as O/Ne properties.If one were to put the Orion Nebula Cluster at a distance of the Galactic Center one would still observe an integrated flux exceeding $10^{-11}$ erg $cm^{-2}$ $s^{-1}$ , at a distance of the LMC $10^{-13}$ erg $cm^{-2}$ $s^{-1}$ . Such long range studies are unprecedented and only possible in the X-ray band between 0.2 and 10 keV with IXO's high spectral resolution capability.